\documentclass[conference]{IEEEtran}
\IEEEoverridecommandlockouts
\usepackage{cite}
\usepackage{amsmath,amssymb,amsfonts}
\usepackage{algorithmic}
\usepackage{graphicx}
\usepackage{textcomp}
\usepackage{xcolor}
\usepackage{amsmath}
\usepackage{subcaption}
\usepackage{xspace}
\usepackage{hyperref}
\usepackage{listings}
\usepackage{booktabs}

\newcommand{\squishlist}{
   \begin{list}{$\bullet$}{%
        \setlength{\itemsep}{0pt}%
        \setlength{\parsep}{0pt}%
        \setlength{\topsep}{0pt}%
        \setlength{\partopsep}{0pt}%
        \setlength{\listparindent}{-2pt}%
        \setlength{\itemindent}{-5pt}%
        \setlength{\leftmargin}{1.2em}%
        \setlength{\labelwidth}{0em}%
        \setlength{\labelsep}{0.5em}%
    }
}
\newcommand{\squishend}{
    \end{list}  }
    
\definecolor{codegreen}{rgb}{0,0.6,0}
\definecolor{codegray}{rgb}{0.5,0.5,0.5}
\definecolor{codepurple}{rgb}{0.58,0,0.82}
\definecolor{backcolour}{rgb}{0.95,0.95,0.92}

\lstdefinestyle{mystyle}{
    backgroundcolor=\color{gray!10!white},   
    commentstyle=\color{codegreen},
    keywordstyle=\color{blue},
    numberstyle=\tiny\color{codegray},
    stringstyle=\color{codepurple},
    basicstyle=\ttfamily\footnotesize,
    breakatwhitespace=false,         
    breaklines=true,                 
    captionpos=b,                    
    keepspaces=true,                 
    numbers=left,                    
    numbersep=5pt,                  
    showspaces=false,                
    showstringspaces=false,
    showtabs=false,                  
    tabsize=2
}

\lstdefinestyle{smaller}{
    backgroundcolor=\color{gray!10!white},   
    commentstyle=\color{codegreen},
    keywordstyle=\color{blue},
    numberstyle=\tiny\color{codegray},
    stringstyle=\color{codepurple},
    basicstyle=\ttfamily\scriptsize,
    breakatwhitespace=false,         
    breaklines=true,                 
    captionpos=b,                    
    keepspaces=true,                 
    numbers=left,                    
    numbersep=5pt,                  
    showspaces=false,                
    showstringspaces=false,
    showtabs=false,                  
    tabsize=2
}
\author{%
\IEEEauthorblockN{%
Zhongming Yu\IEEEauthorrefmark{1},
Genghan Zhang\IEEEauthorrefmark{2},
Hanxian Huang\IEEEauthorrefmark{1},
Xin Chen\IEEEauthorrefmark{3},
Jishen Zhao\IEEEauthorrefmark{1}
}

\IEEEauthorblockA{\IEEEauthorrefmark{1}University of California, San Diego, La Jolla, CA, USA \\
Email: \{zhy025, hah008, jzhao\}@ucsd.edu}
\IEEEauthorblockA{\IEEEauthorrefmark{2}Stanford University, Palo Alto, CA, USA \\
Email: zgh23@stanford.edu}
\IEEEauthorblockA{\IEEEauthorrefmark{3}Intel Corporation, Santa Clara, CA, USA \\
Email: xin.chen@intel.com}
}

\usepackage[linesnumbered,ruled,vlined]{algorithm2e}

\RestyleAlgo{ruled}
\def\BibTeX{{\rm B\kern-.05em{\sc i\kern-.025em b}\kern-.08em
    T\kern-.1667em\lower.7ex\hbox{E}\kern-.125emX}}
\begin{document}

\newcommand{\nickname}{GeoT\xspace}
\newcommand{\todo}[1]{\textcolor{red}{#1}}

\title{GeoT: Tensor Centric Library for Graph Neural Network via Efficient Segment Reduction on GPU}

\maketitle

\begin{abstract}
Graph Neural Networks (GNNs) have recently ignited a surge of innovation, significantly enhancing the processing of geometric data structures such as graphs, point clouds, and meshes. As the domain continues to evolve, a series of frameworks and libraries are being developed to push GNN efficiency to new heights. While graph-centric frameworks have achieved success in the past, the advent of efficient tensor compilers has highlighted the urgent need for tensor-centric frameworks. Yet, efficient tensor-centric frameworks for GNNs remain inadequate due to unique challenges and limitations encountered when implementing segment reduction in GNN contexts: (1) \textbf{Support for well-exploited design space}: The necessity to support variable embedding lengths necessitates moving beyond traditional parallel algorithms, leading to a scarcity of optimized approaches. (2) \textbf{Heuristic adaptability to input dynamics}: The absence of an effective algorithm for parameter selection frequently results in sub-optimal kernel configurations. (3) \textbf{Format agnostic and fusion potential}: Achieving seamless integration within existing frameworks remains an essential yet challenging demand.

In response to these challenges, we introduce \nickname (\underline{Geo}metric \underline{T}ensor), a cutting-edge tensor-centric library designed specifically for GNNs via efficient segment reduction on GPU. \nickname debuts innovative parallel algorithms that not only introduce new design principles but also expand the available design space. It features an input-heuristic kernel selection algorithm that simplifies the optimization process with a low-overhead decision tree. Importantly, \nickname is engineered for straightforward fusion within a computation graph, ensuring compatibility with contemporary tensor-centric machine learning frameworks and compilers.
Setting a new performance benchmark, \nickname marks a considerable advancement by showcasing an average operator speedup of 1.80x and an end-to-end speedup of 1.68x. \footnote{The open-source code and artifact can be found at \url{https://github.com/fishmingyu/GeoT}.}

\end{abstract}

\section{Introduction}

\begin{figure}[!t]
\centerline{\includegraphics[width=0.5\textwidth]{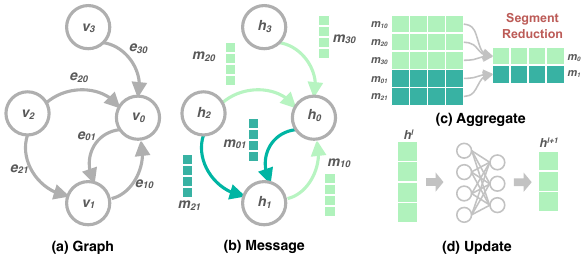}}
    \vspace{-5pt}
    \caption{A simplified illustration of message-passing GNNs schedule. (a) An example graph with 4 nodes and 5 edges. (b) The message stage of GNN corresponded to Eq.~\ref{eq:message} (c) The aggregation stage corresponded to Eq.~\ref{eq:aggregate}. (d) The update step corresponded to Eq. \ref{eq:update}. }
    \vspace{-15pt}
    \label{fig:graphaggr}
\end{figure}

Graph Neural Networks (GNNs)\cite{kipf2016semi, bronstein2017geometric} have significantly broadened the scope of deep learning by extending their application beyond traditional data types like images and videos to geometric structures, including graphs, meshes\cite{verma2018feastnet, pfaff2020learning, siddiqui2023meshgpt}, and point clouds~\cite{qi2017pointnet, qi2017pointnet++}. This extension has enabled GNNs to 
achieve remarkable efficacy across various domains, such as medical research and discovery~\cite{gasteiger2021gemnet, watson2022broadly}, personalized recommendations~\cite{pal2020pinnersage}, traffic forecasting~\cite{jiang2022graph}, and weather prediction~\cite{lam2022graphcast}.
Despite the successes of graph-centric libraries~\cite{wang2019deep}, the emergence of efficient tensor compilers has highlighted the urgent need for tensor-centric libraries~\cite{fey2019fast}. As illustrated in Fig.~\ref{fig:graphaggr}, in tensor-centric libraries, segment reduction is pivotal, representing the aggregation step essential to GNN operations.

Implementing segment reduction in GNNs poses significant challenges due to the implicit segmentation of inputs and the accommodation of variable-length vector inputs (See Section \ref{segment reduction}. Although prior studies~\cite{merrill2015cub, bell2012thrust} 
have addressed segment reduction, they 
primarily focused on scalar inputs, representing merely a fraction of the potential applications in geometric deep learning. We identify three primary 
critical challenges in optimizing segment reduction for GNNs: \textbf{(1) Support for well-exploited design space:} Supporting variable embedding lengths necessitates the development of innovative parallel algorithms, resulting in an under-explored optimization space. Implementations like torch\_scatter~\cite{torch_scatter}, while addressing these requirements, do not fully exploit modern hardware capabilities and lack efficient heuristics for kernel tuning, leading to suboptimal performance.
\textbf{(2) Heuristic adaptability to input dynamics:} The absence of an effective mechanism for kernel parameter selection frequently results in sub-optimal configurations. The complexities of GPU optimization for deep learning, combined with the extensive design space for kernel parameters, pose significant challenges~\cite{dai2022heuristic}.
\textbf{(3) Format agnostic and fusion potential: }Achieving seamless integration within existing frameworks~\cite{fey2019fast}, especially where message and aggregation fusion is feasible, remains a significant, unaddressed challenge in segment reduction.

In response to these challenges, we 
propose \nickname (\underline{Geo}metric \underline{T}ensor), a tensor-centric library specifically designed for GNNs through efficient segment reduction. \nickname contributes novel solutions to the domain:
\begin{itemize}
\item A customized tiling algorithm for segment reduction, coupled with advanced thread workload mapping and GPU-specific optimizations, to tackle the challenges associated with the under-exploited design space.
\item Deployment of a data-driven, ultra-low overhead decision tree for configuration rule selection, offers heuristic adaptability to dynamic input scenarios. This 
innovative method achieves the comparable performance of optimal tuning without the extensive computational cost, streamlining kernel configuration for varying data characteristics.
\item To address the challenges of integrating fusion within GNN frameworks, \nickname offers a provision of format-agnostic fusion support, which guarantees seamless integration with existing computation-graph-level optimizations. 
This feature addresses the challenges of integrating fusion within GNN frameworks, enhancing operational efficiency and compatibility with modern ML systems.
\end{itemize}
\nickname's contributions not only tackle the existing gaps in segment reduction for GNNs but also pave the way for integration for library and compiler for geometric deep learning via efficient segment reduction. 
Compared to the state-of-the-art baseline, \nickname achieves an average speedup of 1.28x for segment reduction, 1.80x for Sparse Matrix-Matrix Multiplication (SpMM), and 1.68x for end-to-end inference tasks.

\section{Background and Motivation}
\label{background}
\subsection{Graph Neural Networks}
In Fig.~\ref{fig:graphaggr}, we show a basic message-passing GNN computation. To be specific, we consider a graph $\mathcal{G}=\mathcal{(V,E)}$ with nodes $v_{i}\in \mathcal{V}$ and edges $e_{ij} \in \mathcal{E}$ we define a graph neural network layer follow the notation from~\cite{gilmer2017neural}.

\begin{gather}
\textit{message}: \mathbf{m}_{ij} = \phi_e(\mathbf{h}_i, \mathbf{h}_j, a_{ij}) \label{eq:message}
\\
\textit{aggregate}:\mathbf{m}_i = \sum_{j\in \mathcal{N}_i} \mathbf{m}_{ij} \label{eq:aggregate}
\\
\textit{update}:\mathbf{h}_i^{l+1}=\phi_h(\mathbf{h}_i^{l}, \mathbf{m}_i) \label{eq:update}
\end{gather}

The message step (Eq.~\ref{eq:message}) in GNNs exhibits considerable variation. In point cloud-related models such as PointNet~\cite{qi2017pointnet}, the messaging involves complex Multilayer Perceptrons (MLPs). Conversely, in models like GCN~\cite{kipf2016semi} and GraphSAGE~\cite{hamilton2017inductive}, the message can be simplified to the node embedding itself, allowing for fusion with the aggregation step. Equivariant Graph Neural Networks (EGNNs)~\cite{satorras2021n} maintain equivariance to rotations by adding additional computational load to the message processing step.

Despite this variability, the workload of the aggregation step remains relatively consistent. As illustrated in Eq. \ref{eq:aggregate}, aggregation involves summing messages from a node's neighbors. Typically, these neighbors are segmented in non-decreasing order, meaning that $j\in \mathcal{N}_i$ is continuous, and $i$ is sorted, enabling this aggregation to function similarly to segment reduction. The nuances of segment reduction will be further explored in Section \ref{segment reduction}.

The messaging and aggregation processes can be integrated into SpMM (Sparse Dense Matrix Multiplication)~\cite{huang2020ge}, particularly when the messaging component is relatively straightforward, as observed in models like GCN. However, at the heart of this fused message aggregation lies segment reduction, which serves as the fundamental building block. This concept is further explored and detailed in Section \ref{seq:fusion}.

\begin{figure}[!t]
\centerline{\includegraphics[width=0.5\textwidth]{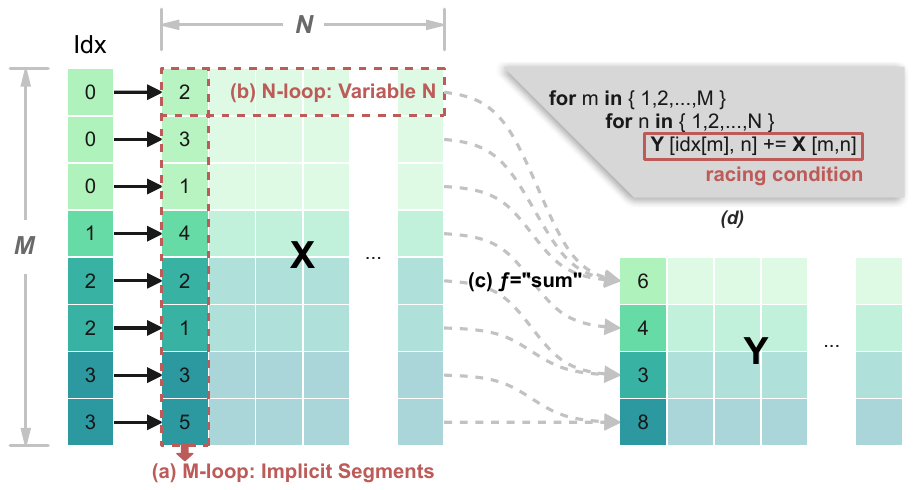}}
    \vspace{-5pt}
    \caption{An illustration of segment reduce operation. (a) M-loop of segment reduction. (b) N-loop of segment reduction. (c) The aggregation's type of reduction $f$ can be varied, such as using mean or max. But the most commonly used aggregation type is sum. (d) A pseudocode of segment reduction. }
    \vspace{-15pt}
    \label{fig:scatter}
\end{figure}

\subsection{Segment Reduction}
\label{segment reduction}

Specifically, Fig.~\ref{fig:scatter} depicts the segment reduction operation. This operation takes two primary inputs: \texttt{Idx}, a 1-D array ordered in a non-decreasing sequence representing the graph's segmented neighbors, and \textbf{X}, symbolizing the edge message matrix used during the aggregation phase. With a feature dimension denoted by $\mathcal{F}$, \textbf{X}'s dimensions are $M \times N$, where $M=\mathcal{|E|}$ and $N=\mathcal{F}$. Segment reduction processes these inputs to produce output \textbf{Y}, with dimensions $\mathcal{|V|} \times N$.

Though segment reduction primarily involves two loops—over M and N—the operation's parallelization is far from straightforward. Firstly, in GNNs, the size of $N$ can range from $1$ to $128$, necessitating different parallelization strategies. Secondly, \texttt{Idx}'s implicit segmentation complicates parallelizing the M-loop due to potential race conditions. An elementary approach to parallelization might employ atomic operations universally, leading to excessive overhead without leveraging the inherent segment characteristics to reduce atomic operations.

Our objective is to devise a superior approach that accommodates the diverse sizes of $N$ while preserving the implicit segmentation through smart and efficient thread management. Detailed discussions of our proposed solutions will follow in subsequent sections.

\begin{figure*}[t]
\centerline{\includegraphics[width=0.98\textwidth]{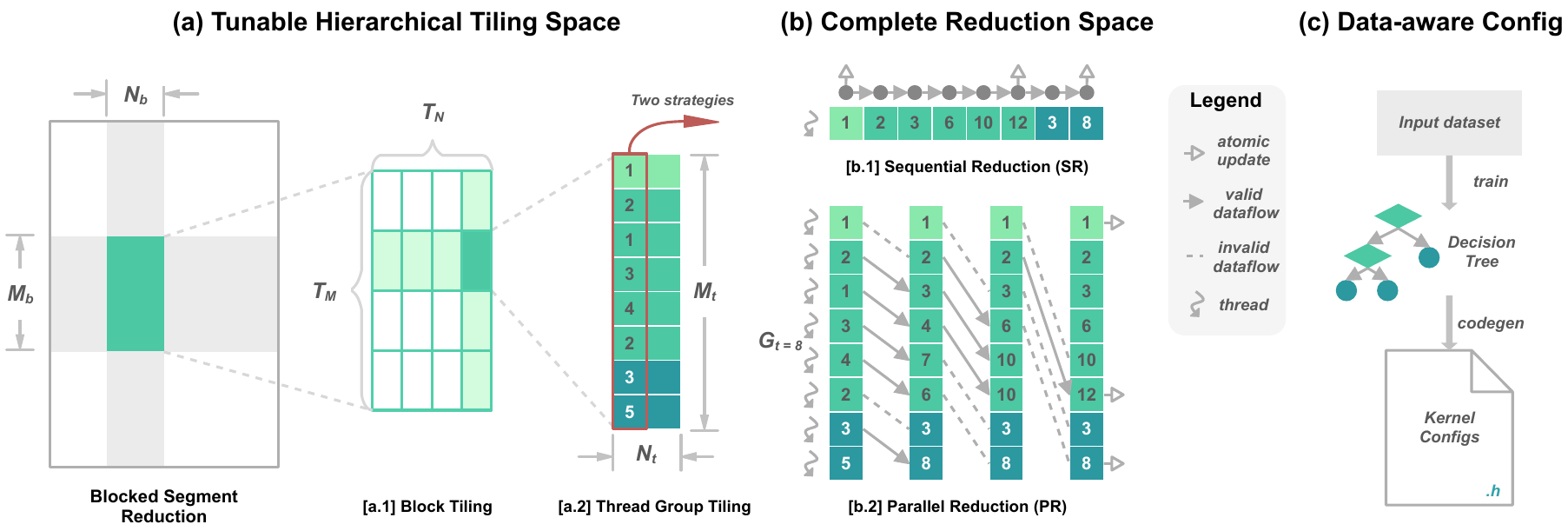}}
    \vspace{-5pt}
    \caption{Design principles for efficient segment reduction include (a) Tunable Hierarchical Tiling Space, utilizing block and thread group tiling for a hierarchical design; (b) Within the thread group's M-loop subtask, employing SR and PR reduction strategies, with PR regulated by a distinct parameter $G_t$ for synchronized thread group management; (c) A decision tree is applied to determine the optimal segment reduction rules, and eventually the code config is automatically generated. }
    \label{fig:principle}
    \vspace{-15pt}
\end{figure*}

\subsection{GNN Frameworks}
The rapid expansion of GNN applications has stimulated the creation of specialized GNN systems and libraries, which largely adhere to one of two design philosophies: graph-centric and tensor-centric frameworks. 

Graph-centric frameworks, which are exemplified by DGL~\cite{wang2019deep}, Neugraph~\cite{ma2019neugraph}, and others~\cite{wang2021gnnadvisor, liu2020g3}, focus on direct manipulations and optimizations of graph data structures. Their operations and workloads are 
designed around graph data structures, utilizing either sparse matrices or constructing graph-specific manipulations.
Yet, these frameworks often encounter difficulties in fully harnessing the capabilities of modern tensor compilers, such as TVM~\cite{chen2018tvm, feng2023tensorir}, XLA~\cite{google_xla}, and Triton~\cite{tillet2019triton}. Such a limitation primarily stems from their dependence on graph-specific data structures and APIs, which are not inherently aligned with the generalized optimizations and functionalities offered by advanced tensor compilers.

In contrast, tensor-centric frameworks, such as PyTorch Geometric (PyG)~\cite{fey2019fast}, offer a more flexible approach by representing graphs and related structures as tensors. Such an abstraction facilitates better integration opportunities with tensor compilers, enhancing performance and scalability. However, such a shift introduces 
substantial challenges 
in optimizing the segment reduction operation, which is a cornerstone in message-passing GNNs.

\subsection{GPUs}
\label{nvgpu}

As demonstrated in MLPerf benchmarks~\cite{reddi2020mlperf}, GPUs have become predominant in the machine learning domain, primarily because of their robust parallel programming capabilities enabled by the Single Instruction, Multiple Threads (SIMT) architecture. Nvidia GPUs, as a prime example, structure their parallel units around Stream Multiprocessors (SMs) and allocate blocks of threads to each SM. Within this architecture, threads are organized into warps, which act as the fundamental unit of SIMT processing~\cite{NVIDIA2023CUDAC}.

CUDA has introduced Cooperative Groups~\cite{nvidia_cooperative_groups} as a method for flexible CUDA Thread Programming. This feature recognizes that the extent of sharing and synchronization needed can vary significantly between different algorithms. Hence, thread synchronization requires flexibility to adapt to these variations. We utilize this tool to facilitate efficient thread management, optimizing our approach to parallel computing within the CUDA environment.

\section{Design Principle}
\label{seq:design}

The design principles of our approach can essentially be divided into two main parts. The first part is dedicated to the parallel algorithm design, featuring our innovative tiling techniques (Section~\ref{seq:2d tiling}) and fine-grained reduction strategies (Section~\ref{seq:reduction}). This portion is fundamental in optimizing parallel algorithms by efficiently partitioning data and computations for enhanced performance.

The second part dives into the utilization of data-aware configuration rules (Section~\ref{seq:data-aware}). These rules are designed to avoid the inefficient process of manual rule design. Instead, by leveraging insights derived from the machine-learning guided rules, our approach aims to achieve performance levels that align with those obtained through traditional tuning processes, but with portability and near-zero overhead.

To better illustrate our design, we list the configurable parameters as shown in Table \ref{tab:config}.

\begin{table}[htbp]
\centering
\caption{Configurable Parameters in Segment Reduction}
\label{tab:config}
\begin{tabular}{lll}
\toprule
Params & Method & Description \\
\midrule
$T_M$ & Block Tiling & Number of thread groups per M-loop \\
$T_N$ & Block Tiling & Number of thread groups per N-loop \\
$M_t$ & Thread Group Tiling & M dimension data per thread group \\
$N_t$ & Thread Group Tiling & N dimension data per thread group \\
$G_t$ & Reduction Space & Synced threads in a thread group \\
\bottomrule
\end{tabular}
\end{table}

Additionally, we address the fusion techniques in a separate, standalone Section~\ref{seq:fusion}.

\subsection{Tunable Hierarchical Tiling Space}
\label{seq:2d tiling}
\textbf{Challenges and Related works:} Hierarchical tiling is a well-established strategy in matrix-oriented parallel computing, frequently applied in tasks such as GEMM (General Matrix Multiplication)\cite{cutlass_cuda} and image processing workloads\cite{ragan2013halide}. One significant limitation these applications face is the lack of potential race conditions, which hinders the adaptability of such techniques to the segment reduction challenge. Optimization efforts for SpMM \cite{yang2018design} have largely focused on format-constrained optimization, leveraging CSR-based approaches\cite{hong2019adaptive, dai2022heuristic} or delving into other sophisticated formats like ELL~\cite{ye2023sparsetir, kjolstad2017tensor}. However, the hierarchical tiling strategies developed in these contexts can't seamlessly align with the needs of segment reduction. Although previous segment reduction libraries~\cite{torch_scatter} have implemented basic tiling approaches, they do not fully exploit the memory coalescence features and the inherent parallelism capabilities of GPUs. In response, we propose a novel blocked segment reduction algorithm that employs hierarchical tiling, thus making the challenge tunable based on future metric selections.

\textbf{Solution:} As depicted in Fig.~\ref{fig:principle}, our approach to blocked segment reduction is structured around two principal hierarchical levels. The first level, block tiling, manages data partition into GPU blocks. Subsequently, the thread group tiling level takes on the role of distributing specific sub-tasks to units within these blocks. In formulating our strategy, we opted not to leverage the shared memory hierarchy, as the implicit segmentation rendered the data ambiguous, making conventional methods such as streaming neighbor embedding via shared memory~\cite{wang2021gnnadvisor, yu2023hypergef} inapplicable to our context.

The partitioning works as follows: for a given data block of dimensions $M_b \times N_b$, which is from matrix \textbf{X}, we assign a block equipped with $T_M \times T_N$ thread groups for its processing. Each thread group is then tasked with handling a $M_t \times N_t$ sub-section of the matrix \textbf{X}. Consequently, we establish that $M_b = T_M \cdot M_t$ and $N_b = N_t \cdot T_N$. Different from previous work~\cite{torch_scatter}, we had new parameters $T_M$ and $N_t$, which enlarges the tunable space for parameter selection.

The metrics $T_M$, $T_N$, $M_t$, and $N_t$ are designated as tunable parameters. This decision allows for a more adaptable and optimized configuration, facilitating the tailoring of our blocked segment reduction algorithm to various scenarios and hardware configurations.

It's important to note that the concept of thread group tiling diverges from the traditional definitions of thread or warp tiling. It can refer to a single thread, a collective of threads (for example, 8 threads), or even up to a full warp (32 threads). The specifics of this distinction will be further detailed in Section~\ref{seq:reduction}.

\subsection{Complete Reduction Space}
\label{seq:reduction}

\textbf{Challenges and Related Work:} 
The challenges of efficient segment reduction derive from the irregular and implicit segmentation in the M-loop and the potential for race conditions within parallel processing units. Effective thread management is essential to minimize atomic operations and comprehensively explore the reduction space.
Although advancements in variable length reduction techniques have been explored \cite{dai2022heuristic, bell2009implementing}, they fall short of establishing a comprehensive reduction space. The thread group technique, initially introduced in Sgap~\cite{zhang2023sgap}, remains constrained by its sparse context and specific format.
Our goal is to create a holistic reduction space framework that allows for precise thread management.

\textbf{Solution:} In the context of the M-loop, there exist two primary types of reduction. The first type, sequential reduction (SR), is visualized in Fig.~\ref{fig:principle} [b.1], where a singular thread handles the reduction, atomically updating the output solely at a segment's ending. Conversely, parallel reduction (PR) leverages multiple threads to carry out the reduction task simultaneously. Regarding the N-loop, both SR and PR approaches process it successively to harness Instruction Level Parallelism (ILP), ensuring atomic updates are restricted to the end of a segment, as illustrated in Fig.~\ref{fig:principle}.

\SetKwComment{Comment}{\textcolor{blue}{// }}{}

{\scriptsize
\begin{algorithm}
\caption{Segment Reduction With PR}\label{alg:PR}
\DontPrintSemicolon
\KwData{Idx, $G_t$, $M_t$, $\mathbf{Y, X}$, $\mathcal{F}$}
\KwResult{Modified $\mathbf{Y}$ based on segment Idx and $\mathbf{X}$}

$n_G \gets M_t/G_t$\;
$nid \gets \_cal\_nid()$\;
\Comment{\textcolor{blue}{Calculate segment}}
$is\_seg \leftarrow 
((group.shfl\_up(key, 1) \neq key) \lor (group.thread\_rank() == 0))$\;
\Comment{\textcolor{blue}{Tmp registers}}
Declare ValueType $tmpv$, $o[n_G]$\;
Declare IndexType $tmpr$\;
\Comment{\textcolor{blue}{Group shfl down}}
\For{$i \leftarrow 0$ \KwTo $n_G$}{
  $mid \gets \_cal\_mid(n_G)$\;
  $o[i] \gets \mathbf{X}[mid \times \mathcal{F}+nid]$\;
  $key \gets Idx[mid]$\;
  \For{$k \leftarrow 1$ \KwTo $G_t$ \textbf{using} $k \leftarrow k \ll 1$}{
    $tmpv \leftarrow group.shfl\_down(o[i], k)$\;
    $tmpr \leftarrow group.shfl\_down(key, k)$\;
    \If{$tmpr == key$ \textbf{and} $group.thread\_rank() < (G_t - k)$}{
      $o[i] \leftarrow o[i] + tmpv$\;
    }
  }
  \Comment{\textcolor{blue}{Atomic update to Y}}
    \If{$is\_seg$}{
    $atomicAdd(\mathbf{Y} + key \times \mathcal{F}+nid, o[i])$\;
}
}

\end{algorithm}
}

To implement PR effectively, we introduce $G_t$, a parameter indicating the number of synced threads within the thread group. This parameter is explicitly set above one to differentiate it from SR, yet it does not exceed $M_t$ to avert out-of-bounds issues. Typically, $G_t$ is adjusted to a power of 2, aligning with GPU architectural preferences. For instance, a scenario with $G_t=8$ is depicted in Fig.~\ref{fig:principle} [b.2]. To elucidate the PR reduction space, the algorithm for predetermined $M_t$ and $G_t$ values is detailed below. As outlined in Algorithm~\ref{alg:PR}, the initial step involves segment identification for determining the relevant sections. Following this, the group shuffle-down algorithm is employed. This parallel reduction examines the fine-grained details of each segment through \texttt{tmpv}, ensuring that dataflow between segments is correctly invalidated, as shown in Fig.~\ref{fig:principle}. This process guarantees accurate accumulation. Ultimately, PR culminates in an atomic update to \textbf{Y}, informed by the insights derived from the segments previously analyzed.

By integrating both SR and PR strategies and introducing $G_t$ as a novel tunable parameter for PR, we effectively expand the reduction space. This comprehensive approach accommodates the full algorithmic spectrum of segment reduction, enhancing overall efficiency and adaptability.

\begin{figure}[ht]
\vspace{-5pt}
  \centering
  \begin{subfigure}[b]{0.26\textwidth}    
    \includegraphics[width=\textwidth]{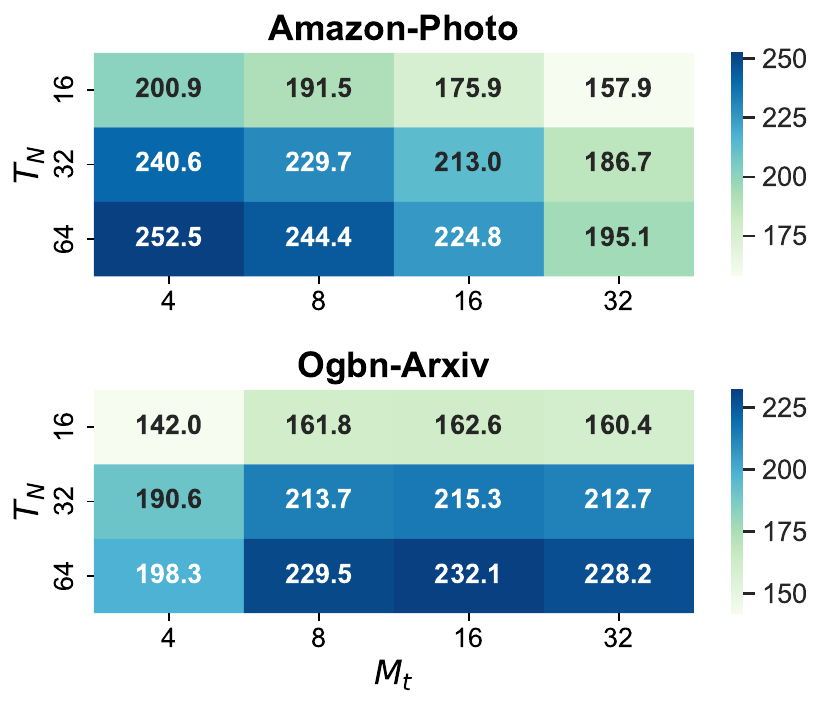}    
    \caption{Config Heatmap}
    \label{fig:heatmap}
  \end{subfigure}
  \hfill    
  \begin{subfigure}[b]{0.22\textwidth}    
    \includegraphics[width=\textwidth]{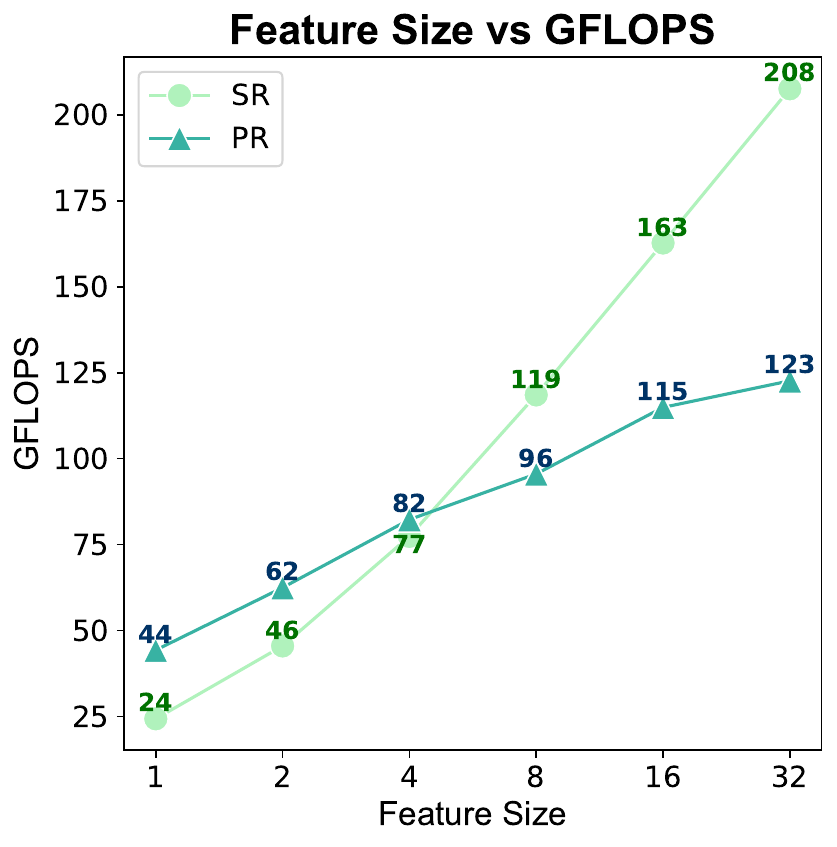}
    \caption{SR and PR comparison}
    \label{fig:avg_gflops}
  \end{subfigure}
  \caption{(a) The GFlops heatmaps for the datasets Amazon-Photo and Ogbn-Arxiv highlight the impact of varying configuration settings on performance. Specifically, two configurations, $T_N$ and $M_t$, are chosen to illustrate how different config settings can influence computational efficiency.
(b) The representation of average GFlops across augmented datasets about SR and PR methods provides insights into the potential trade-offs between these two reduction strategies.}
  \label{fig:main}
\vspace{-18pt}
\end{figure}

\subsection{Data-aware Config Rules}
\label{seq:data-aware}
\textbf{Motivation:} In reality, datasets show 
significant variation. As illustrated 
in Fig.~\ref{fig:heatmap}, different datasets distinctly require
different configurations. For instance, the optimal configuration for the Amazon-Photo dataset is markedly different from that for Ogbn-Arxiv. Even within a single dataset, varying parameter selections can lead to substantial performance disparities, with potential speedups of up to 1.6x between the lowest and highest performance. It should be emphasized that although our demonstration is limited to just two datasets and a selection of configuration dimensions, the challenge of data-aware configurations is remarkably pronounced across other datasets and configurations.

Hence, identifying the appropriate configuration for a given input presents a notable challenge. Specifically, for the segment reduction problem in GNNs, with given input \texttt{Idx} and $\mathcal{F}$, we aim to determine an optimal configuration set $<T_N, T_M, M_t, N_t, G_t>$ and scheduling strategy (SR or PR) to achieve the best performance without a large overhead.

\textbf{Challenges and Related work:} Choosing the right configurable parameters, as outlined in Table \ref{tab:config}, is a considerable challenge owing to the intricate design space they create. These parameters are deeply interrelated, with each affecting the others and collectively impacting the overall performance and efficiency of the system. Moreover, the extensive range of parameter options contributes to an exponential increase in complexity.
Conventional optimization libraries for DNNs~\cite{cuda,onednn}, GNNs~\cite{torch_scatter, huang2020ge}, and arithmetic~\cite{cublas,oneMKL} often rely on hand-crafted heuristics and accumulated experiential knowledge for determining optimization parameters. This traditional approach often results in sub-optimal performance due to its static nature. These heuristics are not universally effective across different datasets, tasks, or evolving hardware architectures, leading to a lack of adaptability and inefficiencies in computational resource usage. Furthermore, the manual tuning process is both time-consuming and expertise-dependent, making it challenging to scale and adapt to new problems or data. 

\begin{figure}[t]
\centerline{\includegraphics[width=0.48\textwidth]{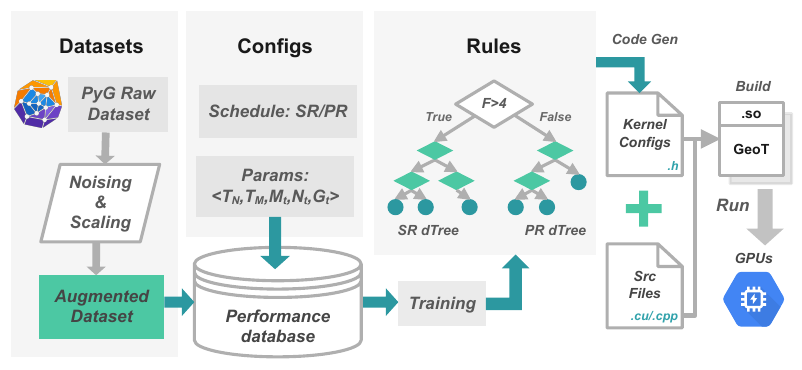}}
    \vspace{-0pt}
    \caption{The process flow of data-aware configuration. The key idea of this part is to utilize a performance database to train more efficient decision-tree rules for segment reduction. The decision tree is transformed into kernel configs through code generation, facilitating the compilation of the final .so library for GPUs. }
    \label{fig:data_config}
    \vspace{-18pt}
\end{figure}

Moving beyond traditional techniques, the application of machine learning (ML) to learn optimization strategies has gained traction within DNN compilers and frameworks. This approach is exemplified by innovations such as AutoTVM~\cite{chen2018learning}, Ansor~\cite{zheng2020ansor}, Flextensor~\cite{flextensor}, Chameleon~\cite{chameleon}, Halide~\cite{Adams19}, and One-shot Tuner~\cite{oneShotTuner}. The realm of GNN workload research, too, has seen ML-based rule configurations being employed, a notable example being dgSPARSE~\cite{dai2022heuristic}. Through training on datasets comprised of optimal schedules, ML-based methodologies can discern the essential heuristics for optimal parameter determination. While these ML-driven solutions offer adaptability to specific task characteristics, resulting in precise parameter selection and improved performance, their application within libraries presents challenges. The main challenge arises from the runtime implications of ML-based rule configurations. While this approach is feasible for just-in-time (JIT) tensor compilers, it presents a substantial hurdle for tensor libraries. The integration of these ML methods incurs substantial overhead and poses implementation difficulties within pre-built libraries.

\begin{figure*}[!ht]
    \begin{minipage}[t]{0.48\textwidth}
\begin{lstlisting}[language=Python, caption=PyG fusion with sparse format, label={lst:pyg_sparse}]
import torch_sparse
from torch_sparse import SparseTensor

# prepare sparse tensor(usually in CSR)
adj = SparseTensor(row=edge_index[0], col=edge_index[1], values=None)
# transpose
adj_t = adj.t()
...

# Aggregate messages based on SpMM
out = torch_sparse.matmul(adj_t, x, reduce="sum")
\end{lstlisting}
    \end{minipage}\hfill
    \begin{minipage}[t]{0.48\textwidth}
        \begin{lstlisting}[language=Python]
import geot

# use index operation for message
msg = x[edge_index[0]]
# use segment_reduce for aggregation
out = geot.segment_reduce(edge_index[1], msg, reduce="sum")
\end{lstlisting}
        \begin{lstlisting}[language=Python, caption=GeoT solution for fusion, label={lst:geot}]
# Aggregate messages based on fused segment reduction
out = geot.index_segment_reduce(edge_index[0], edge_index[1], x, reduce="sum")
\end{lstlisting}
\end{minipage}
\vspace{-25pt}
\end{figure*}

\textbf{Solution:} To address the challenges highlighted, our approach begins with rule-based parameter pruning to streamline the design space. This initial step is followed by the adoption of a data-driven machine learning (ML) method to efficiently and effectively derive configuration rules, moving away from the labor-intensive process of manual rule determination.

To narrow down the vast design space, we refine the range of candidate parameters based on parallel programming in the field~\cite{huang2020ge, zhang2023sgap}. The choices for $T_N$ are constrained to $\{16, 32, 64\}$, and for $T_M$ to $\{1, 2, 4, 8\}$. We set the range for $M_t$ as $\{8, 16, 32, 64\}$, for $N_t$ as $\{1, 2, 4\}$, and for $G_t$ as $\{8, 16, 32\}$ (For SR we ignore $G_t$, since it is equal to 1). By implementing these prunings, we effectively reduce the complexity to constant space, making the configuration process more manageable.

For the process of the ML approach, as illustrated in Fig.~\ref{fig:data_config}, we introduce a process for data-aware configuration. Our primary technique is to deploy a decision tree~\cite{quinlan1986induction} to determine the most appropriate configuration accurately. The decision tree stands out as a lightweight ML technique, introducing negligible overhead since its decision nodes can be transformed into \texttt{if-else} statements, which incur only nanoseconds of cost on contemporary processors~\cite{intel_core_i9_13900k}. Contrary to the conventional implementation of the decision tree rule-based library outlined by Thomas et al.~\cite{thomas2005framework}, which typically involves mutually exclusive leaf nodes, our approach utilizes a multi-output decision tree regressor~\cite{scikit_learn_tree_regression}. This method allows for the selection of the configuration set in its entirety, considering the correspondence of parameters rather than treating them as isolated entities. 

For training the decision tree, a genuine performance dataset is essential. Initially, we gather 51 valid datasets from the PyG~\cite{fey2019fast} dataset collection and employ noising and scaling techniques to expand this to 3060 datasets. Subsequently, we explore a comprehensive range of configurations, encompassing both schedules and parameters, and benchmark the outcomes offline on GPUs.

By adopting the concept of a performance database~\cite{thomas2005framework}, we construct a database where the key comprises the dataset's features, schedule, and parameters, as demonstrated in Fig.~\ref{fig:data_config}, and the value represents the GFlops performance of the corresponding kernel. Notably, we opt for $O(1)$ complexity features, such as the size of \texttt{Idx} ($Idx\_size$), the maximum number of \texttt{Idx} ($Idx\_max$), and the average ($avg=Idx\_size/Idx\_max$). Given that \texttt{Idx} is inherently sorted, determining the maximum feature incurs $O(1)$ complexity. This strategy further minimizes the feature extraction overhead during the configuration selection phase.

The training dataset for the decision tree is sourced from the performance database, employing the Top 1 selection rule for data retrieval. The features of the input data, specifically $Idx\_size$, and $avg$, plus feature size $\mathcal{F}$ serve as the decision nodes within the tree, while the configuration parameters $<T_N, T_M, M_t, N_t, G_t>$ are designated as leaf nodes. It's important to note that the choice between SR (Sequential Reduction) and PR (Parallel Reduction) schedules is made based on empirical analysis. This decision is informed by the observation that the trade-off between SR and PR is significantly influenced by the size of the features involved. Particularly for PR, as depicted in Fig.~\ref{fig:principle} [b.2], the threads operate sequentially along the M-loop. Given the row-major organization of the \textbf{X} matrix, an increase in $\mathcal{F}$ can lead to a memory access pattern that is not coalesced, resulting in inefficiencies\cite{fauzia2015characterizing}. Therefore, as illustrated in Fig.~\ref{fig:avg_gflops}, based on our empirical findings, we opt for the SR (Sequential Reduction) schedule when $\mathcal{F}>4$, and reserve the remaining scenarios for PR (Parallel Reduction). Furthermore, we maintain two distinct decision trees to cater to SR and PR schedules, respectively.

Armed with these rules, we proceed to generate code that forms a kernel config file. This file, along with other source files, is then compiled to construct the \nickname .so library.

\section{Format-agnostic Fusion}
\label{seq:fusion}
\textbf{Motivation:} Fusion plays a critical role in modern ML systems~\cite{google_xla,wu2023pytorch,niu2021dnnfusion}, as it significantly reduces DRAM memory footprint and minimizes redundant memory access by retaining tensors on the chip. We aim to introduce a format-agnostic fusion method that facilitates seamless integration of the aggregate step (segment reduction)  and message step in GNNs.

\textbf{Challenges and Related Work:} Format-agnostic fusion often goes overlooked in GNN systems, which typically emphasize format-specific fusion. However, the principle of format agnosticism is widely recognized in tensor-centric ML systems, especially within modern end-to-end ML compilers~\cite{ding2023hidet, lai2023relax,zhao2023canvas}. Format agnosticism means the input tensor is treated as dense, eliminating the need for specific formats such as ragged tensors~\cite{fegade2022cora} or sparse tensors~\cite{gale2020sparse}. Despite the availability of sparse tensor compilers like TACO~\cite{kjolstad2017tensor} and SparseTIR~\cite{ye2023sparsetir}, their integration into mainstream end-to-end frameworks remains challenging. This issue derives from the fact that nodes within the computation graph are format agnostic, thereby leading to a large gap between DNN and GNN systems due to their differing approaches to fusion.

One reason for this gap is that many GNN systems adhere to graph processing paradigms~\cite{wang2016gunrock, liu2020g3}, positioning the graph as the central element of the system. Compilers such as Graphiler~\cite{xie2022graphiler}, HGL~\cite{gui2022hgl}, SeaStar~\cite{wu2021seastar}, and PIGEON~\cite{wu2023pigeon} are developed atop graph-centric systems, rendering them domain-specific and thereby limited.

Moreover, even without an explicit focus on graphs, previous endeavors still stick with representing graphs in sparse formats. Several works~\cite{chen2020fusegnn, rahman2021fusedmm, zhang2022understanding} concentrate on fusing techniques for sparse matrices, particularly in CSR format, while others like~\cite{wang2021gnnadvisor, huang2021understanding} strive to optimize the CSR format through neighbor grouping strategies.

\textbf{Solution}: To overcome these challenges, we have developed a method that seamlessly combines the aggregation and messaging phases through our advanced segment reduction technique. This method integrates all strategies outlined in \ref{seq:design}, including the application of decision tree rules. To clarify our approach, we present code samples in Listing~\ref{lst:pyg_sparse} and Listing~\ref{lst:geot}, which simplify the graph convolution process detailed by Morris et al.\cite{morris2019weisfeiler}, focusing on messaging and aggregation. The upper part of  Listing~\ref{lst:geot} demonstrates how \nickname performs without fusion using \texttt{segment\_reduce}. In this process, $edge\_index[1]$, serving as the $Idx$ for segment reduction, is kept in a non-decreasing sequence, as guaranteed by GNN frameworks~\cite{fey2019fast}. The message phase involves indexing the input feature $x$ and conveying the $msg$ to the segment reduction, identified as \textbf{X} in \ref{segment reduction}. Consequently, indexing and \texttt{segment\_reduce} operate the same operation shown in Listing~\ref{lst:pyg_sparse}, albeit without fusion.

To merge the message and aggregation phases, Listing~\ref{lst:geot} introduces a novel \texttt{index\_segment\_reduce} function, an enhancement over \texttt{segment\_reduce} that requires only slight code adjustments. Likewise, for incorporating weight considerations in SpMM, we developed a \texttt{index\_weight\_segment\_reduce} function. These functions demand minimal code alterations, leveraging the tiling, reduction techniques, and decision rules formulated for \texttt{segment\_reduce}. The distinction between the fusion API and the \texttt{segment\_reduce} data-configuration process (as illustrated in Fig.~\ref{fig:data_config}) lies in our approach to modifying the kernel config function API without altering the underlying rules. Consequently, we compile a new .so library, incorporating the newly introduced fusion functions. 

Contrast with the sparse format fusion method used in PyG (Listing~\ref{lst:pyg_sparse}), which depends on torch\_sparse's~\cite{pytorch_sparse} SpMM in a sparse tensor format (usually CSR), our GeoT approach delivers even faster results with greater flexibility (See Section~\ref{exp:fusion}. Additionally, the format-agnostic fusion approach developed by \nickname offers the potential for seamless integration into compilers in the future, such as Triton~\cite{tillet2019triton}. This capability underscores \nickname's versatility and forward compatibility with emerging technologies in the domain of machine learning compilers, paving the way for broader applicability and optimization opportunities across various ML frameworks.

What's more, by eliminating the prerequisites for format configuration and sparse tensor development, our solution is particularly beneficial for GNN systems that utilize sampling~\cite{bai2021efficient, yang2022gnnlab, serafini2021scalable}. This adaptability is crucial for environments where new graphs are frequently generated.
\section{Evaluation}
\subsection{Implementation}
\label{exp:impl}
For a fair evaluation, we developed \nickname as a torch-extension library built upon PyTorch~\cite{paszke2019pytorch}. The extension includes CUDA~\cite{cuda} code, specifically utilizing the cooperative group feature~\cite{nvidia_cooperative_groups} for the Parallel Reduction (PR) schedule. Decision tree training employs sklearn~\cite{pedregosa2011scikit}, a widely used machine learning toolkit. To facilitate end-to-end inference, our library integrates seamlessly with PyG~\cite{fey2019fast}.

\subsection{Experiment Setup}
\subsubsection{Devices}
\begin{itemize}
    \item \textbf{Nvidia A100}~\cite{nvidia_a100}: Powered by the Ampere architecture, this 80GB PCIE model boasts 1.94TB/s HBM2e bandwidth and 108 SM cores, operating on the CUDA v12.1 platform. Unless specified otherwise, the A100 GPU serves as the standard for our experiments.
    \item \textbf{Nvidia H100}~\cite{nvidia_h100}: With the Hopper architecture, this 80GB PCIE model provides 2.04TB/s HBM2e bandwidth and 112 SM cores. It utilizes the CUDA v12.1 platform and is specifically used in our portability tests.
    \item \textbf{Nvidia RTX 3090Ti}~\cite{nvidia_rtx_3090}: Based on the Ampere architecture, this 24 GB PCIE card features 1.01TB/s GDDR6 bandwidth and 84 SM cores, supported by the CUDA v11.8 platform. This device is also designated for portability experiments.
\end{itemize}

\begin{figure*}[t]
\centerline{\includegraphics[width=0.98\textwidth]{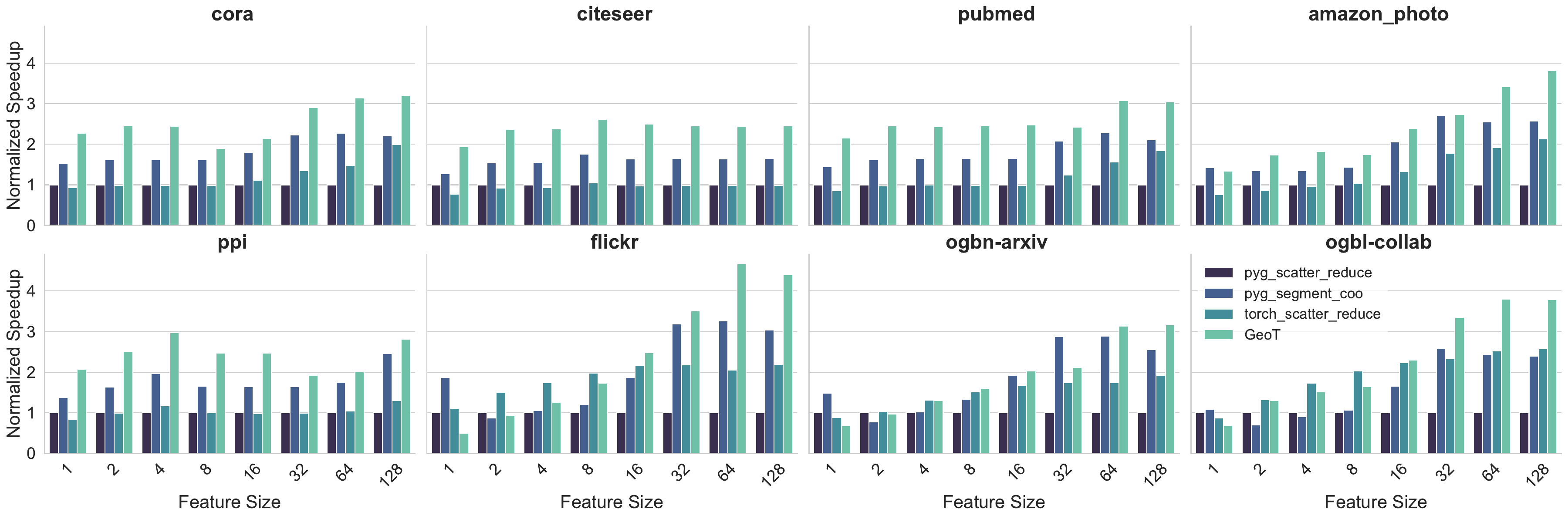}}
    \vspace{-5pt}
    \caption{Overall performance of \nickname's segment reduction compared to SOTA baselines. Performance is normalized based on PyG's \texttt{scatter\_reduce}~\cite{torch_scatter}. The Reddit dataset is excluded since it will lead to the OOM problem. } 
    \label{fig:index_scatter}
\end{figure*}

\subsubsection{Evaluation Datasets}
We selected well-known graph datasets from PyG~\cite{fey2019fast} and OGB~\cite{hu2020open} for our evaluation, as detailed in Table \ref{tab:dataset_stats}. For the Reddit dataset, we opted for a sparser version~\cite{wu2019simplifying, zeng2019graphsaint} due to the original dataset's inability to fit within a single system for end-to-end testing. Nevertheless, the dataset's extensive number of edges could lead to out-of-memory (OOM) issues during segment reduction testing, given the direct correlation between the size of input \textbf{X} and the number of edges. Thus, we didn't test Reddit in segment reduction experiments.

\begin{table}[htbp]
\vspace{-5pt}
\centering
\caption{Evaluation Datasets}
\label{tab:dataset_stats}
\begin{tabular}{lrrr}
\toprule
Dataset & Nodes & Edges &  avg degree \\
\midrule
Citeseer~\cite{bojchevski2017deep} & 3,327 & 9,104 & 2.74 \\
Cora~\cite{bojchevski2017deep} & 2,708 & 10,556 & 3.90 \\
PPI\cite{zitnik2017predicting} & 2,245 & 61,318 & 27.31 \\
Pubmed~\cite{bojchevski2017deep} & 19,717 & 88,648 & 4.50 \\
Amazon-Photo\cite{shchur2018pitfalls} & 7,650 & 238,162 & 31.13 \\
Flickr~\cite{zeng2019graphsaint} & 89,250 & 899,756 & 10.08 \\
Ogbn-Arxiv~\cite{wang2020microsoft} & 169,343 & 1,166,243 & 6.89 \\
Ogbl-Collab~\cite{wang2020microsoft} & 235,868 & 1,285,465 & 5.45 \\
Reddit2~\cite{zeng2019graphsaint} & 232,965 & 23,213,838 & 99.65 \\
\bottomrule
\end{tabular}
\end{table}

\subsubsection{Evaluation Models}
\begin{itemize}
    \item \textbf{GCN}~\cite{kipf2016semi} The Graph Convolutional Network (GCN), is one of the first and most fundamental GNN models. It represents a significant milestone in graph representation learning, proposing an efficient layer-wise propagation rule that leverages the graph structure and node features to generate node embeddings. 
    \item \textbf{GIN}~\cite{xu2018powerful} Expanding the capabilities of GNNs, the Graph Isomorphism Network (GIN), introduces a more powerful framework designed to capture the discriminative power necessary for graph isomorphism problems. GIN is based on the insight that the representational capacity of a GNN is crucial for distinguishing between different graph structures.
    \item \textbf{GraphSAGE}~\cite{hamilton2017inductive} Addressing the scalability and inductive learning challenges in graph neural networks, GraphSAGE is designed for inductive learning, allowing it to generate embeddings for unseen nodes after the model has been trained. 
\end{itemize}
\subsubsection{Baselines}
\squishlist
    \item \textbf{Segment Reduction:} We employ torch\_scatter~\cite{torch_scatter}, an extension library managed by the PyG team~\cite{fey2019fast}, renowned for its highly optimized scatter and segment operations. Specifically, we utilize \texttt{scatter\_reduce} and \texttt{segment\_coo} functions, both yielding identical outputs for segment reduction tasks. Notably, \texttt{segment\_coo} can leverage the sorting of $Idx$ for performance gains. We include \texttt{scatter\_reduce} in our comparison as it outperforms \texttt{segment\_coo} under certain conditions, making it a vital consideration alongside Torch's~\cite{paszke2019pytorch} fast \texttt{scatter\_reduce} implementation.

    \item \textbf{Fusion (SpMM):} For the fusion of messaging and aggregation into SpMM operations, cuSPARSE~\cite{naumov2010cusparse} serves as a primary benchmark. We utilize Torch's library for its convenient Python API that interfaces with cuSPARSE's kernels. Additionally, we compare against pytorch\_sparse~\cite{pytorch_sparse}, another extension library from the PyG team designed for optimized sparse matrix operations.

    \item \textbf{End-to-end Inference:} Given that our development builds upon PyG, our end-to-end inference comparison excludes DGL~\cite{wang2019deep} and similar frameworks to avoid discrepancies in system design and implementation specifics. PyG operates in two modes: Dense mode and Sparse mode. The Sparse mode employs efficient aggregation via SpMM for fusion~\cite{pytorch_geometric_sparse_tensor}, aligning with our evaluation framework.
\squishend

This comprehensive set of baselines encompasses both specific operation optimizations and the broader context of end-to-end system performance, ensuring a thorough assessment of \nickname's effectiveness in segment reduction and fusion tasks.

\subsection{Segment Reduction}
\subsubsection{Overall Results}

Fig.~\ref{fig:index_scatter} presents the comprehensive outcomes of our segment reduction performance evaluation. We utilize PyG's \texttt{scatter\_reduce} as the benchmark for normalization. In comparison with PyG's \texttt{segment\_coo}, \nickname achieves a 1.28x geomean speedup, with a maximum speedup of 1.85x. Additionally, when compared to Torch's \texttt{scatter\_reduce}, \nickname demonstrates a 1.68x geomean speedup over \texttt{scatter\_reduce}, with a maximum speedup of 4.67x. Also, \nickname performs well in parallel scalability, considering our solution remains powerful when $\mathcal{F}$ grows larger. 
These results underscore \nickname's enhanced efficiency in executing segment reduction tasks, reflecting the effectiveness of its optimization strategies against the established benchmarks within the PyTorch Geometric and Torch environments.

While \nickname exhibits remarkable performance gains on average, we observed outliers, particularly with datasets such as Flickr, Ogbn-Arxiv, and Ogbn-Collab, when the feature size $\mathcal{F}=1$. This underperformance relative to the baseline in scenarios of minimal feature size can be ascribed to two primary factors:

\begin{figure*}[t]
\centerline{\includegraphics[width=0.98\textwidth]{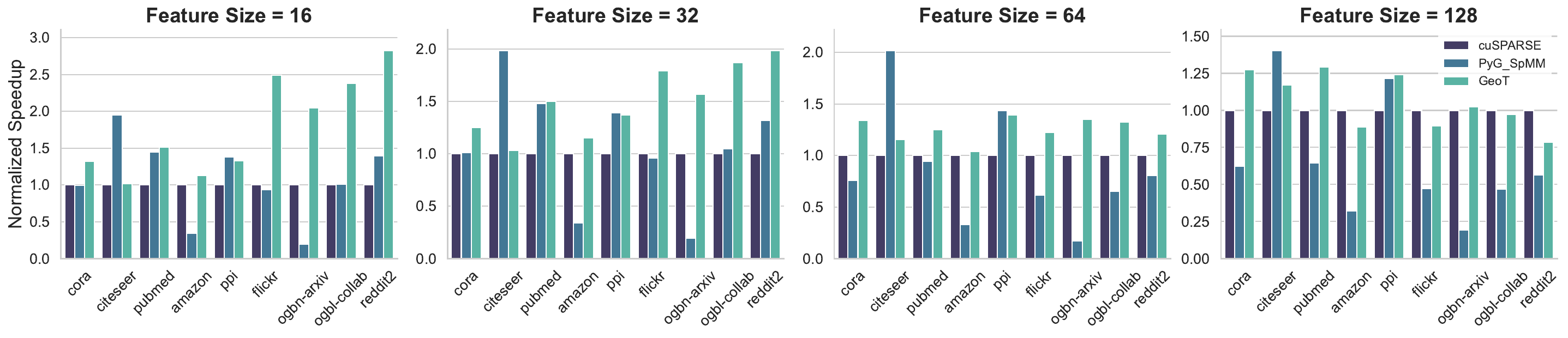}}
    \vspace{-5pt}
    \caption{Overall performance of \nickname's SpMM compared to SOTA baselines. Performance is normalized based on cuSPARSE. }
    \label{fig:spmm}
    \vspace{-10pt}
\end{figure*}

Firstly, for smaller $\mathcal{F}$ values, the Parallel Reduction (PR) schedule is preferred. However, PR entails five configurable parameters, which complicates the convergence process more than the Sequential Reduction (SR) schedule, which has four configurations due to $G_t=1$. Consequently, the decision tree designated for PR configurations tends to be less effective than that for SR, leading to the derivation of suboptimal rules.

Secondly, the case of $\mathcal{F}=1$ introduces additional overhead in managing the Cooperative Group. Unlike \nickname, the baseline implementation in PyG's \texttt{segment\_coo} employs a distinct scheduling approach for this specific scenario, one that does not rely on Cooperative Group management. This specialized scheduling in the baseline allows for more efficient handling of such cases, highlighting an area for potential improvement in \nickname's approach.

\begin{figure}[ht]
    \vspace{-10pt}
    \centering
    \begin{minipage}[t]{0.218\textwidth}
    \vspace{0pt}
        \begin{lstlisting}[style=smaller, language=C++, caption={A brief example of generated decision tree}, label=lst:dtree]
// <Nt, TN, Mt, TM>
// Gt=1 for SR
if (avg <= 1.15) {
    segreduce_sr<1, 32, 4, 8>(idx, X, Y);
} else {
    if (idx_size <= 6413.5) {
        segreduce_sr<1, 32, 4, 8>(idx, X, Y);
    } else {
        segreduce_sr<1, 64, 8, 4>(idx, X, Y);
    }
}
\end{lstlisting}
    \end{minipage}
    \hfill 
    \begin{minipage}[t]{0.242\textwidth}
     \vspace{0pt}
        \centering
        \includegraphics[width=\textwidth]{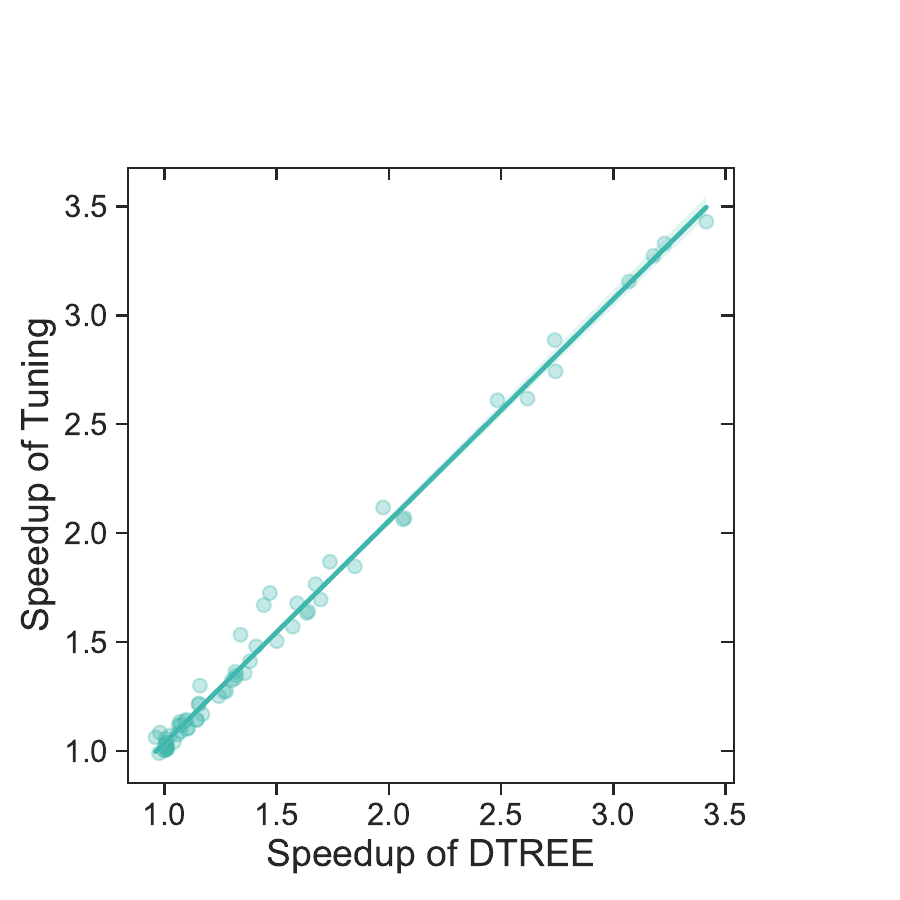} 
        \captionof{figure}{Decision tree and best tuning comparison}
        \label{fig:dtree}
    \end{minipage}
    \vspace{-10pt}
\end{figure}

\subsubsection{Ablation study of decision tree}
In Section \ref{seq:data-aware}, we detail the training process of the multi-output decision tree, specifically setting the maximum depth to 5 to mitigate overfitting. In the generated kernel configs from codegen, we incorporate 31 Sequential Reduction (SR) leaf nodes and 31 Parallel Reduction (PR) leaf nodes, all delineated under specific decision nodes. A snippet of the generated SR code, given that $G_t=1$ which reduces the configuration to four parameters $<N_t, T_N, M_t, T_M>$, is presented in Listing~\ref{lst:dtree}. The decision-making criteria encompass $avg$, $idx\_size$, and $\mathcal{F}$, as elucidated in Section~\ref{seq:data-aware}.

To assess the impact of decision tree optimization, we conducted experiments, the results of which are displayed in Fig.~\ref{fig:dtree}. The comparison baseline is established through a hand-crafted, rule-based segment reduction approach where configurations, as outlined in Table~\ref{tab:config}, are determined based on engineering experience. For instance, the configuration $T_N$ is adjusted concerning $\mathcal{F}$, while $T_M$ correlates with $avg$. The outcomes of our study, spanning various features and datasets, demonstrate that decision tree-derived rules not only outperform conventional hand-crafted approaches but also closely match the optimal performance achieved through extensive tuning.

For the fused version of segment reduction, as detailed in Section~\ref{seq:fusion}, we maintain the established rules but modify the function to either \texttt{index\_segment\_reduce} or \texttt{index\_weight\_segment\_reduce}. This modification is feasible as it only necessitates minor code adjustments to our segment reduction process, such as incorporating indexing. Optimizing the fusion workload could refine this approach by fine-tuning the rules specifically for fused operations.

\subsection{Fusion}
\label{exp:fusion}
In \nickname, Sparse Matrix-Matrix Multiplication (SpMM) is modeled as a fusion of indexing (message passing) and segment reduction (aggregation). To ensure a fair comparison, we employ \texttt{index\_weight\_segment\_reduce} for our evaluation, mirroring cuSPARSE's requirement for specifying sparse tensor values. Fig.~\ref{fig:spmm} displays SpMM performance across feature sizes of $\mathcal{F}=16, 32, 64, 128$.

We benchmark against cuSPARSE's SpMM for normalization purposes. Compared to cuSPARSE, \nickname achieves a geometric mean speedup of 1.34x, with a peak speedup of 2.83x. Notably, when benchmarked against PyG's SpMM implementation, \nickname shows a 1.80x geometric mean speedup, with a remarkable maximum speedup of 10.47x. Plus our results demonstrates commendable extensibility across various datasets.

However, \nickname's fusion approach encounters limitations with large feature sizes ($\mathcal{F}$). Despite strategies to minimize atomic operations through sophisticated thread management, as illustrated in Fig.~\ref{fig:principle}, atomic bandwidth limitations may still pose a bottleneck. Nevertheless, in practical GNN models like GCN~\cite{kipf2016semi} and GIN~\cite{xu2018powerful}, feature sizes $\mathcal{F}$ commonly range from 16 to 64. Analysis by Gu et al.~\cite{gu2021principled} suggests that the effective hidden embedding size $\mathcal{F}$ seldom exceeds 100.

Therefore, \nickname presents a promising approach to fusion, offering a faster SpMM operator without concerns over tensor formats, well-suited to the common feature size ranges found in real-world GNN applications.

\begin{figure}[t]
\centerline{\includegraphics[width=0.48\textwidth]{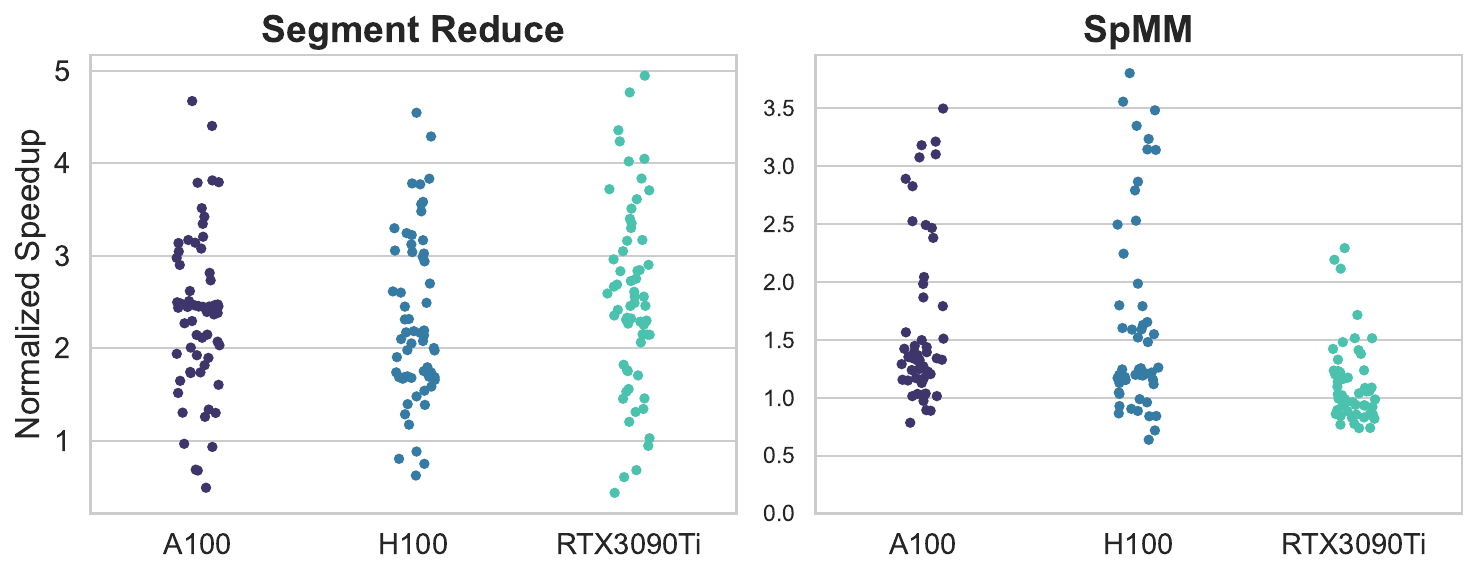}}
    \vspace{-5pt}
    \caption{Overall speedup across various feature sizes and datasets is evaluated in experiments conducted on A100, H100, and RTX 3090Ti GPUs. In the context of segment reduction, PyG's \texttt{scatter\_reduce} serves as the normalization baseline. Similarly, for SpMM operations, cuSPARSE is utilized as the normalized baseline.}
    \label{fig:portability}
    \vspace{-15pt}
\end{figure}

\subsection{Portability}
To evaluate portability, experiments were conducted on H100 and RTX 3090Ti GPUs, with the performance database and decision tree rules initially derived from A100 GPU benchmarks. As depicted in Fig.~\ref{fig:portability}, \nickname demonstrates commendable portability in the segment reduction operator, maintaining consistent speedup across different GPU architectures. For the SpMM operator, the performance portability extends to the H100 GPU; However, a performance drop is observed on the RTX 3090Ti, primarily due to its GDDR6 bandwidth being lower than the HBM2e memory bandwidth found in the H100 and A100 GPUs. Despite this, a comparable geometric mean speedup of 1.08x over cuSPARSE is still achieved on the RTX 3090Ti.

\begin{figure}[ht]
  \vspace{-5pt}
  \centerline{
    \includegraphics[width=0.48\textwidth]{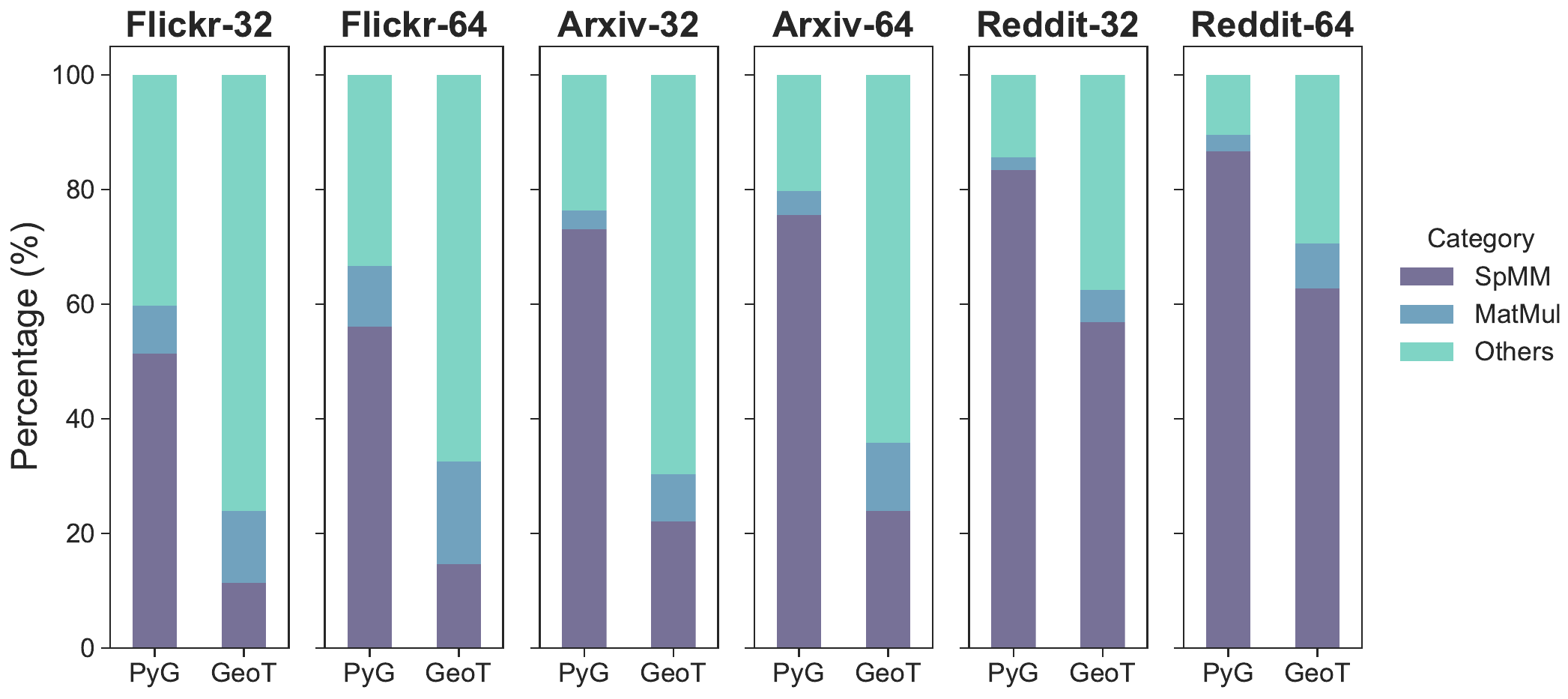}}
  \caption{CUDA time breakdown analysis for GCN model}\label{fig:breakdown}
  \vspace{-10pt}
\end{figure}

\subsection{End-to-end Models}
In practical model implementation, the message and aggregate steps are typically fused into a single operator, often represented by SpMM. We apply the techniques described in Section~\ref{seq:fusion} and integrate them into the end-to-end model. Our testing of \nickname is focused mainly on the inference setup, as designing the associated autograd function for training is part of our future work.

For our end-to-end task, we select node classification, setting the number of layers to three across GCN, GIN, and GraphSAGE models. To more effectively illustrate the ratio of SpMM in the performance breakdown, we use large graphs such as Flickr, Ogbn-Arxiv, and Reddit2 (Ogbl-Collab is utilized for link prediction). The hidden size $\mathcal{F}$ is set to 32 and 64, and the number of output classes is aligned with the classes present in the data.

\subsubsection{Breakdown}
As illustrated in Fig.~\ref{fig:breakdown}, which benchmarks the GCN model, \nickname significantly decreases the percentage of time spent on SpMM operations. On average, there is a 39.1\% reduction in the SpMM portion, reaching up to a maximum reduction of 51.6\%. These detailed breakdowns are consistent with the overall results observed.
\subsubsection{Overall Results}
\begin{figure}[t]
\centerline{\includegraphics[width=0.48\textwidth]{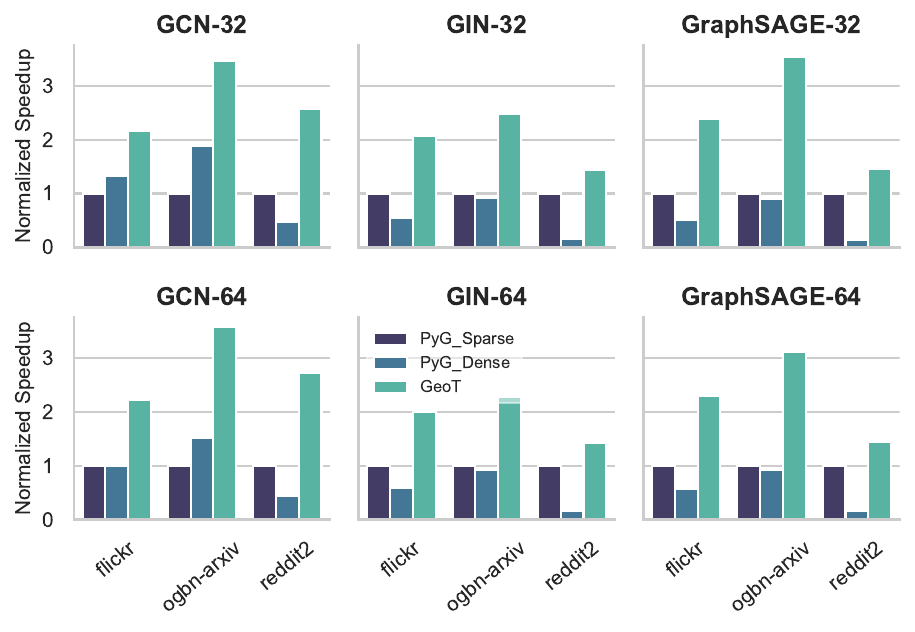}}
    \vspace{-5pt}
    \caption{Design principle of efficient segment reduction. (a) Tunable 2D-Tiling Space. We leverage block tiling and thread group tiling to achieve 
    }
    \label{fig:end2end}
    \vspace{-10pt}
\end{figure}

As depicted in Fig.~\ref{fig:end2end}, \nickname outperforms the baseline across all models. The average speedup of \nickname over PyG's sparse implementation is 1.68x, with a maximum speedup of 3.56x. Against PyG's dense implementation, \nickname achieves an average speedup of 2.36x, with a maximum speedup of 3.56x. The variance observed in the Reddit dataset performance across models primarily stems from GCN's use of SpMM with weight, whereas other models do not incorporate a weight attribute for edges.

The experiments demonstrate \nickname's advantages not only in the single segment reduction operation but also in its performance in fusion (SpMM), achieving satisfactory end-to-end performance. Additionally, \nickname is adaptable to various hardware platforms, further illustrating its versatility and effectiveness.

\section{Discussion and Future works}
\textbf{Diverse Reduction Types:}
As illustrated in Fig.~\ref{fig:scatter}, the range of reduction types extends beyond mere summation. However, we note that generalizing the reduction operation constitutes mainly an engineering effort and does not influence the scheduling of our proposed algorithms.

\textbf{Training and Autograd Support:}
For the autograd mechanism of segment reduction and the fused SpMM operator, additional operations, such as SDDMM~\cite{nisa2018sampled, yu2021exploiting}, are implicated. Notably, our approach to format-agnostic fusion proves particularly advantageous for training phases, as it hinders the need to maintain a transposed CSC sparse format during backpropagation~\cite{huang2021understanding}, simplifying the overall process.

\textbf{Fine-Tuning Configuration Rules:} While our experiments demonstrate satisfactory results for fusion and portability, the current rules are derived from a performance database specific to the A100 GPU, focused on segment reduction. For enhanced performance, establishing a new performance database and fine-tuning the configuration rules based on fusion workloads and various GPUs would be advantageous.

\section{Conclusion}

In this paper, we introduce \nickname, a tensor-centric library for GNNs via efficient segment reduction. We have innovatively developed a hierarchical tiling space and a comprehensive reduction space algorithm to facilitate parameter tuning. To address the complex configuration rules challenge, we propose a method based on the multi-output decision tree, enabling low-overhead and efficient data-aware configurations. Additionally, \nickname supports seamless fusion without the need for format specification, laying the groundwork for future compatibility with systems and compilers. Supported by extensive experiments, \nickname surpasses all baselines in average benchmarks, showcasing its capabilities not just at the operator level but also in end-to-end model implementations.
\section{Acknowledgement}
We extend our gratitude to Mingfei Ma and Ding Ke from Intel Corporation for their invaluable support, feedback, and suggestions. This work has been partially supported by NSF grants 1829524, 1817077, and 2011212, as well as the PRISM center within JUMP 2.0, an SRC program sponsored by DARPA. The opinions, findings, conclusions, or recommendations expressed in this material are solely those of the authors and do not necessarily reflect the views of the National Science Foundation.
\bibliographystyle{IEEEtran}
\bibliography{ref}

\end{document}